\documentclass[structabstract]{aa}  
\usepackage{graphicx}
\usepackage{txfonts}

\begin{document}

\title{
Constraints on the radial distribution of the dust properties in the CQ~Tau protoplanetary disk
}
\author{F. Trotta \inst{1,2,3,4} \and L. Testi\inst{3,4,5} \and A. Natta\inst{3,6} \and A. Isella\inst{7} \and L. Ricci\inst{7}}
\institute{
Department of Physics and Astronomy, University of Bologna, viale Berti Pichat 6/2, 40127 Bologna, Italy \\
\email{francesco.trotta4@unibo.it}
\and
Dipartimento di Fisica e Astrofisica, Universit{\'a} degli Studi di Firenze, Largo E. Fermi 2, 50125 Firenze, Italy
\and
INAF - Osservatorio astrofisico di Arcetri, Largo E. Fermi 5, 50125 Firenze, Italy
\and
ESO, Karl Schwarzschild str. 2, D-85748 Garching bei Muenchen, Germany 
\and
Excellence Cluster Universe, Boltzmannstr. 2, D-85748, Garching, Germany
\and
Dublin Institute for Advanced Studies, School of Cosmic Physics, 31 Fitzwilliam Place, Dublin 2, Ireland
\and
Division of Physics, Mathematics and Astronomy, California Institute of Technology, MC 249-17, Pasadena, CA 91125, USA
}
\date{Received ... / Accepted ...}

\authorrunning{Trotta et al.} %
\titlerunning{Constraining variations in dust properties in CQ Tau disk}
\abstract 
{Grain growth in protoplanetary disks is the first step towards the formation of the rocky cores of planets.
Models predict that grains grow, migrate, and fragment in the disk and predict varying dust properties as a
function of radius, age, and physical properties. High-angular resolution observations at more than one (sub-)mm wavelength are the essential tool for constraining grain growth and migration on the disk midplane.}
{We developed a procedure to analyze self-consistently multi wavelength (sub-)mm continuum interferometric
observations of protoplanetary disks to constrain the radial distribution of dust properties.}
{We apply this technique to existing multi frequency continuum mm observations of the disk around CQ Tau, a
 A8 pre-main sequence star with a well-studied disk.}
{We  demonstrate that our models can be used to simultaneously constrain 
the disk and dust structure. In CQ Tau, the best-fitting model has a radial dependence of the maximum grain size, which 
decreases  from a few cm in the inner disk ($< 40$ AU) to a few mm at 80 AU. Nevertheless, the currently available 
dataset does not allow us to exclude the possibility of a uniform grain size distribution at a 3$\sigma$ level.
}
{}
\maketitle

\section{Introduction}
\label{sec:intro}

Planetary systems are relatively common around all types of stars (e.g. Borucki et al. 2011).
The origin of planets is intricately tied to the evolution of their primordial
protoplanetary disks. The disks provide the reservoirs of raw material and determine the conditions
for the formation of  planetary systems.
A possible evolutionary scenario currently considered for planet formation is the 
\textit{core accretion gas-capture} model (see e.g. Lissauer 1993), featuring a build-up
of planet cores first and then a gas-infall on sufficiently massive cores later.
According to this model, the growth by coagulation from the initial ISM population of sub-micron size dust grains
to centimeter sizes is  the first step of planet formation.

Spatially resolved observations at sub-(mm) wavelengths of the thermal dust emission from circumstellar disks
provide the most direct tool for studying dust properties
and thus investigating the very early stages of planet formation.
Except for the  inner disk region, the dust optical depth at sub-(mm) wavelengths is sufficiently low, so that
it is possible at these wavelengths to directly probe   the dust grain populations near the disk midplane,
where the process of planet formation is supposed to occur.

In the past two decades, several authors have measured the spectral energy 
distribution (SED) of protoplanetary disks at (sub)-mm wavelengths and found 
that it is possible to describe it as a power-law with index $\alpha$  ($F_{\nu} \propto \nu^{\alpha}$),
with $\alpha$ being significantly lower in this case than in the diffuse interstellar medium case, where $\alpha \sim 3.5 - 4$
(e.g. Beckwith \& Sargent 1991, Wilner et al. 2000, Testi et al. 2001, 2003,
Natta et al. 2004, Wilner et al. 2005, Andrews \& Williams 2005, Rodmann et al. 2006, 
Lommen et al. 2007, 2009, Ricci et al. 2010a, b, 2011).  Assuming optically thin 
emission and a Rayleigh-Jeans approximation, $\alpha$ is directly linked to the spectral 
index $\beta$ of the dust opacity coefficient, where $k_\nu \propto \nu^{\beta}$ and 
$\beta=\alpha-2$. In this way, measured low values of $\alpha$ ($\lesssim 3$) translate 
into a value of $\beta \lesssim 1$ which can be explained if grains have grown to sizes of at 
least a few millimeters (see e.g. Draine 2006, Natta et al. 2007).

More recently, spatially resolved, multi wavelength interferometric observations have been carried 
out for a few disks as a means to study the radial dependence of grain growth (Isella et al. 2010, Banzatti et al. 2011, 
Guilloteau et al. 2011, Perez et al. 2012). The results indicate a radial dependence 
of the spectral index $\beta$, which appears to increase from low values (in some cases, as low as 0) in 
the inner disk to about 1.7--2 in the outer disk regions. This is interpreted as a decrease in the maximum 
grain size as a function of the distance from the cetral star, from centimeter-size grains in the inner disk to mm-size and smaller grains in the outer disk 
(typically, $\sim 100$ AU).
Most of these results are based on the method presented in Isella et al. (2010). 
The method requires spatially resolved observations at two (or more) (sub)-mm wavelengths, 
which are independently compared to theoretical disk models to constrain the disk 
surface density and temperature by assuming a constant dust opacity, or a constant $\beta$ through the disk.
The best-fit models at the different wavelengths are then compared to 
derive the true radial profile of $\beta$ and of the grain size distribution and 
the disk surface density and dust temperature (see, e.g., Per\'ez et al. 2012). 

We propose to constrain the disk structure and the radial variation in the grain size 
distribution through a more direct approach. We develop a self-consistent disk model that includes 
a grain size distribution that varies with the distance from the central star. The model is used to 
investigate the radial dependence of the mm-wave spectral indexes $\alpha$ and $\beta$ from the 
variations in the maximum grain size and to calculate the radial profile of the disk emission to compare 
with the observations.

We apply the model to analyze a set of interferometric observations of the pre-main sequence star 
CQ Tau. The star CQ Tau is one of the closest (100 pc, Hipparchos, Perryman et al. 1997), young intermediate mass stars
surrounded by a circumstellar disk. It is situated in the Taurus-Auriga star forming region and is a well-studied 
variable star of the UX Ori class with spectral type A8 and an estimated age $\sim 5 - 10$ Myr 
 (e.g. Chapillon et al. 2008). The dust continuum emission from the CQ Tau disk has been studied in detail 
 by several authors in the past. The presence of grain growth to $\sim$ cm sizes was inferred from 
 the analysis of the disk SED at mm-wavelengths (Natta et al. 2000, Chiang et al. 2001, Testi et al. 2001,2003, 
 Chapillon et al 2008).
A multi frequency study of CQ Tau was done by Banzatti et al. (2011), who analyzed high-resolution observations
from $0.87$ mm to $7$ mm obtained at the SMA, IRAM-PdB, and NRAO-VLA interferometers.
Assuming a constant dust opacity throughout the disk, they found evidence of evolved dust
by constraining the disk-averaged dust opacity spectral index $\beta$ to be 0.$6 \pm 0.1$.
Different surface density profiles are found at different wavelengths,  indicating a change of dust 
properties with radius.They concluded suggesting a radial variation in the slope $\beta$ of the dust opacity, 
which is consistent with a degree of grain growth varying across the disk.

In this paper, we simultaneously analyze  the three datasets at $1.3$, $2.7$ and $7$ mm used in Banzatti et al. (2011),
relax the assumption of radially constant dust opacity and fit the data with a disk model, where the 
grain size distribution is allowed to vary with the distance to the central star.

The paper is organized as follows:
Section~\ref{sec:disk_model} and \ref{sec:radial_var} contains the description of the adopted disk structure 
and dust opacity models and some examples of the results. The procedure used to fit the data is described 
in Section~\ref{sec:fitting}. In Section~\ref{sec:results} and \ref{sec:discussion}, we present the results 
and discuss the possible origin of the fitted maximum grain size radial distribution.
The main conclusion of this study are summarized in Section~\ref{sec:conclusions}.

\section{Disk model}
\label{sec:disk_model}

The gas surface density profile $\Sigma_g(R)$ is described
by the similarity solution for the disk surface density of a viscous Keplerian disk 
(Lynden-Bell \& Pringle 1974). We adopt the mathematical  formulation 
described in Isella et al. (2009):  
\begin {equation}
   \Sigma_g(R,t) = \Sigma_{tr} \left( \frac{R_{tr}}{R} \right)^{\gamma}
    \exp \left\{ -\frac{1}{2(2-\gamma)} \left[ \left( \frac{R}{R_{tr}} \right)^{(2-\gamma)} - 1 \right] \right\},
    \label{eq:selfsimilar_solution} 
\end {equation}
where $R_{tr}$ is the transition radius\footnote{radius at which the mass flow through the disk 
changes direction, so that the resulting mass flow for $R<R_{tr}$  is directed inward (disk accretion) 
and outward for $R > R_{tr}$  (disk expansion).}, $\Sigma_{tr}$ is the surface density at $R_{tr}$, 
and $\gamma$ is the power law index of the viscosity $\nu_t(R) \propto R^{\gamma}$.
We assume a constant dust-to-gas mass ratio $\zeta=0.01$ throughout the disk, so that dust surface 
density $\Sigma_d$ is $100$ less massive than the gas surface density $\Sigma_g$ at each radius $R$.

The  physical structure and emission of the disk is calculated by adopting the two-layer approximation 
for the solution of the radiation transfer equation (Chiang \& Goldreich, 1997), 
following the procedure presented in Dullemond et al. (2001). The  disk is described 
through a warm surface layer directly heated by the stellar radiation 
and a cooler midplane interior heated by the radiation reprocessed by the surface layer. 
Both temperatures are calculated as a function of the orbital radius by assuming that  the disk is in 
hydrostatic equilibrium between the gas pressure and the stellar gravity. This leads to a flared 
geometry with the opening angle that increases with the distance from the central star.
The central star is treated as a blackbody. For CQ Tau, we assume the same stellar properties found
 in Testi et al. (2003):  $T_{\rm{eff}} = 6900$ K, $L_\star = 6.6~L_\odot$ $M_\star = 1.5~M_\odot$, and $d=100$ pc.

In the disk interior, we adopt a dust model as found in Banzatti et al. (2011), which consists of porous composite spherical grains 
made of 7\% astronomical silicates, 21\% carbonaceous material, 42\% water ice, and 30\% vacuum.
We use the Bruggeman effective medium theory (Bruggeman 1935) to calculate an effective 
dielettric function $\epsilon_{\rm{eff}}$ for the composite grain and then use this function in the Mie 
theory to derive the dust absorption efficiency $Q_{\rm{abs}}(a,\nu)$. The complex optical constant
of individual components are taken from Weingartner \& Draine (2001) for astronomical silicates, from 
Zubko et al.(1996) for carbonaceous material, and from Warren (1984) for water ice.
Throughout the entire disk, we consider a grain size distribution parametrized as a truncated power-law with slope $q$:
\begin {equation}
   n(a) \propto a^{-q} \hspace{0.5cm} \textsl{with} \hspace{0.5cm}  a_{\rm{min}} < a < a_{\rm{max}}
   \label{eq:grainsizedistr_PL} 
\end {equation}
truncated between the minimum and maximum grain sizes $a_{\rm{min}}$ and $a_{\rm{max}}$, respectively.
The value of $a_{\rm{min}}$ is set to 5~nm, but the results of our analysis do not depend on the exact value of 
this parameter as long as $a_{\rm{min}} << 1$ mm. For the power law index, we have used $q=3$ as proposed 
in previous studies and theoretical expectations (Draine 2006).

Unlike previous works, we allow the maximum grain size $a_{\rm{max}}$ to vary with the 
orbital radius. We adopt a power law profile:
\begin {equation}
   a_{\rm{max}}(R) = a_{\rm{max,0}} \left( \frac{R}{R_0} \right)^{b_{\rm{max}}},
   \label{eq:amax_PL} 
\end {equation}
where $a_{\rm{max,0}}$ is the maximum grain size value at a reference position $R_{0}$, and
$b_{\rm{max}}$ is the power-law exponent.
Given the dust absorption efficiency $Q_{\rm{abs}}(a,\nu)$ and the grain size distribution $n(a,R)$,
the dust opacity at frequency $\nu$ and radius R can be then expressed as  
\begin {equation}
   k_\nu^i(R) = \zeta \times \frac{\pi \int n(a,R) a^2 Q_{\rm{abs}}(a,\nu) da}{(4 \pi / 3) \rho_s \int n(a,R) a^3 da},
\end {equation}
where $\rho_s$ is the mean density of composite dust grain given in our case, such that
 $\rho_s = f_{\rm{sil}} \rho_{\rm{sil}} + f_{\rm{ca}} \rho_{\rm{ca}} + f_{\rm{ice}} \rho_{\rm{ice}}$, 
where the $f_i$ terms are the volume fractions of each species given above.

In the disk atmosphere, we assume a constant grain size distribution characterized by grains 
smaller than 1 $\mu$m. This choice, which does not affect the radial profile of the 
millimeter-wave disk emission,  is motivated because larger grains are thought 
to settle toward the disk midplane (see, e.g., Dullemond \& Dominik 2004).

The disk emission is finally expressed as the sum of the emission from the disk 
interior $I^i_\nu(R)$ and the disk surface layer $I^s_\nu(R)$.  
These are expressed (see Dullemond et al. 2001), respectively, by 
\begin{equation}
   I_{\nu}^i(R) = \frac{ \cos{i}}{d^2}  B_\nu(T_i(R))
               \left[ 1 - e^{- \frac{\Sigma(R) k^i_\nu(R)}{\cos{i}}} \right]
\end{equation}

\begin{equation}
  I_{\nu}^s(R) = \frac{1}{d^2} B_\nu(T_s(R)) \Delta \Sigma(R) k_\nu^s(R)
               \left[ 1 + e^{- \frac{\Sigma(R) k^i_\nu(R)}{\cos{i}}} \right],
\end{equation}
where $d$ is the distance to the source, $i$ is the disk inclination defined as the 
angle between the disk axis and the line of sight to the disk ($i=0$ means face-on), and $\Delta\Sigma$ is the
column density in the disk surface.

\section{Effect of the radial variation in the maximum grain size on  disk emission}
\label{sec:radial_var}

\begin{figure}[]
   \centering
   \includegraphics[width=7.5cm]{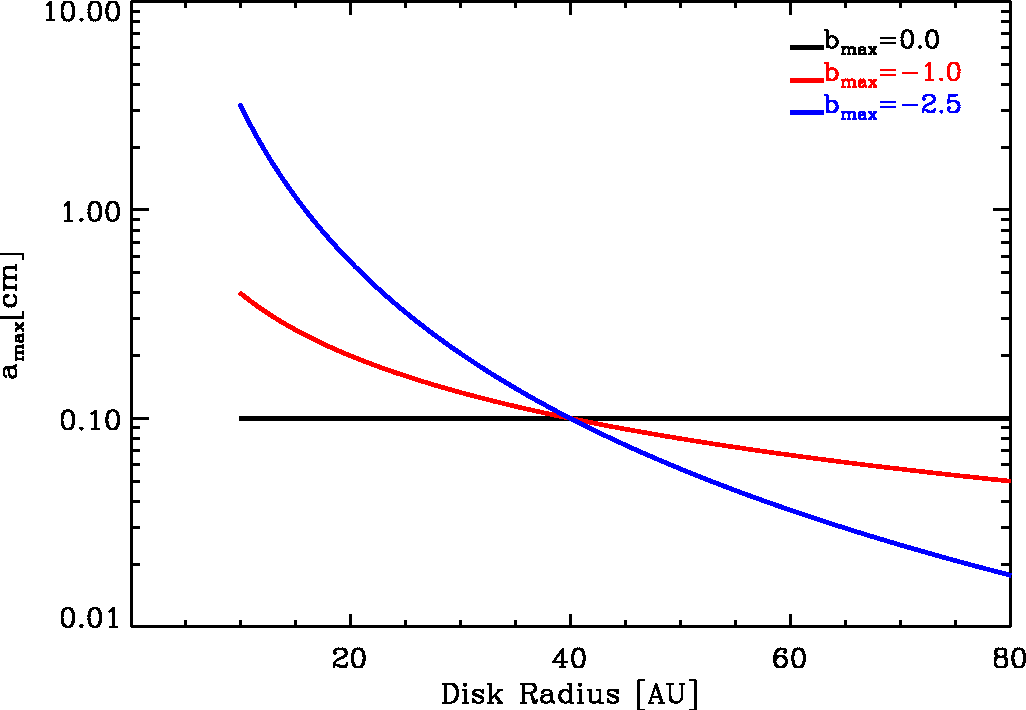}
   \caption{Radial profile of the maximum grain size $a_{\rm{max}}(R)$.
            Different curves represent different values of the parameter $b_{\rm{max}}$ as labeled but have the same $a_{\rm{max,0}}=1$ mm at 40 AU.}
   \label{fig:amax_radius}
\end{figure}

\begin{figure}[]
   \centering
   \includegraphics[width=7.5cm]{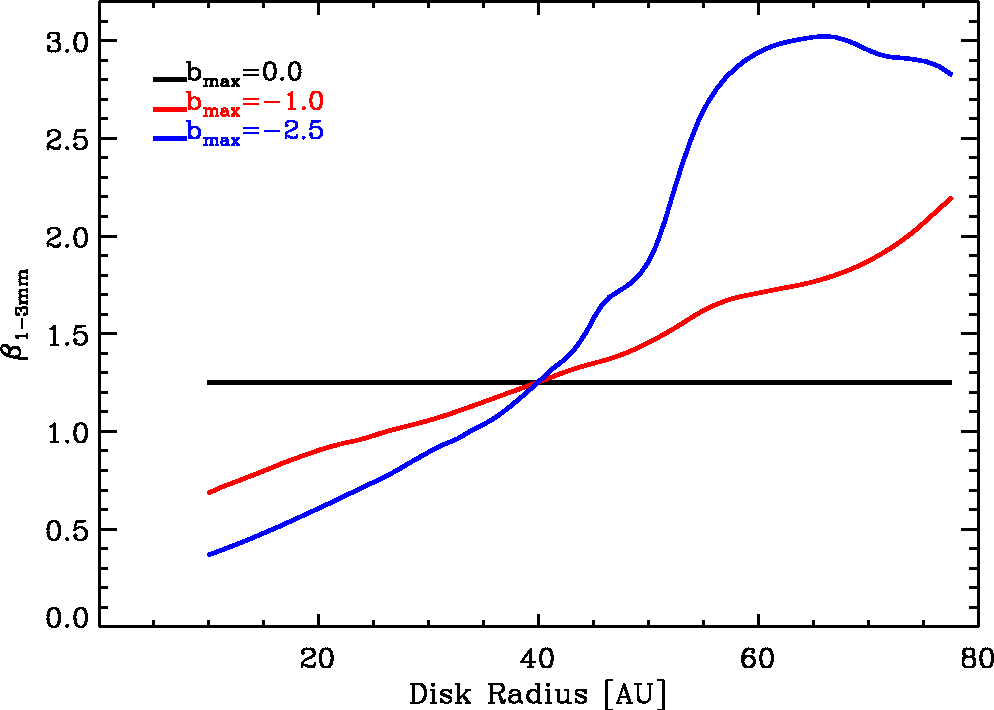}
   \caption{Radial profile of $\beta_{\rm{1-3mm}}(R)$.
            Different curves represent different values of the parameter $b_{\rm{max}}$ as labeled but have the same $a_{\rm{max,0}}=1$ mm at 40 AU.}
   \label{fig:beta_radius}
\end{figure}

\begin{figure}[]
   \centering
   \includegraphics[width=7.5cm]{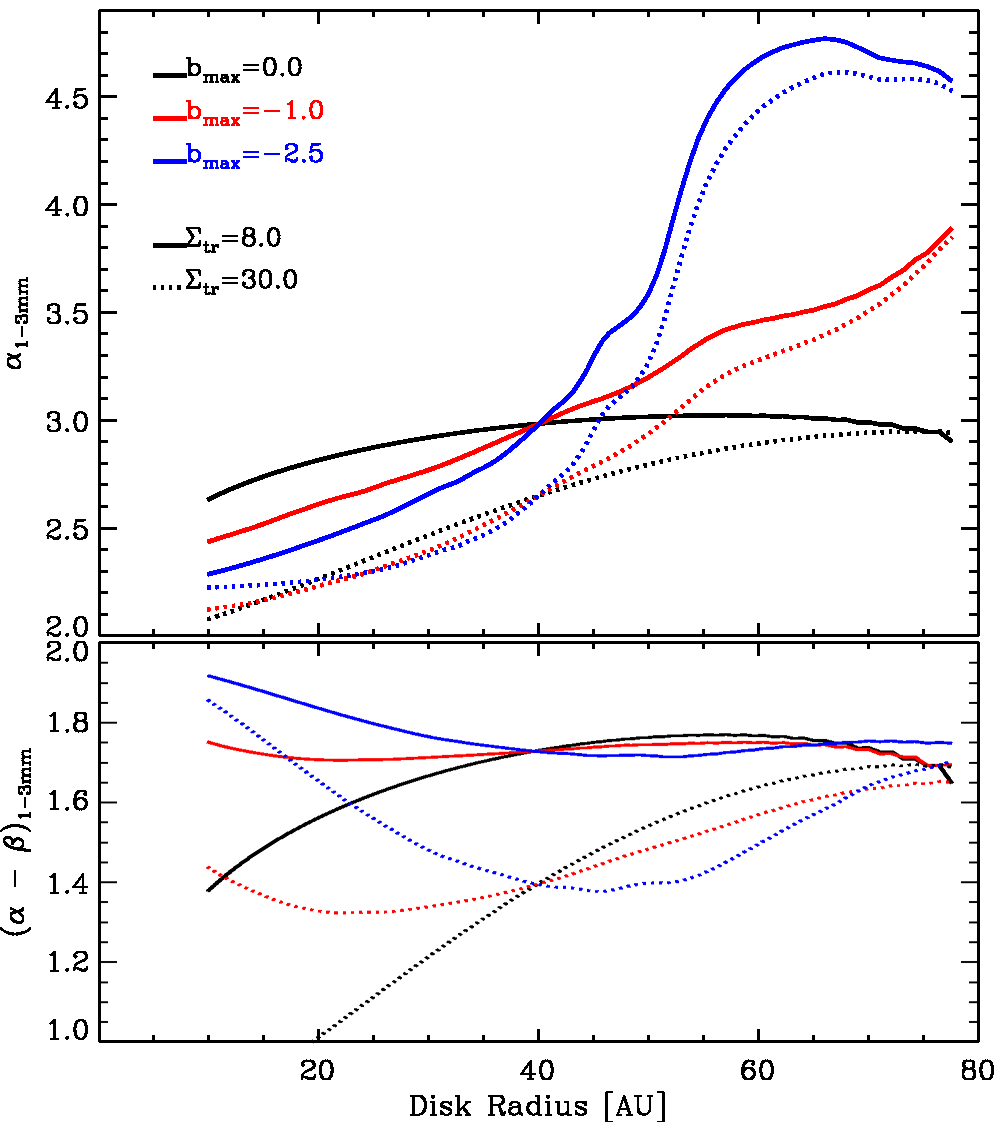}
   \caption{Radial profiles of the spectral index $\alpha$ between $1$ and $3$ mm.
            Different curves represent different values of the parameter $b_{\rm{max}}$ as labeled but have the same $a_{\rm{max,0}}=1$ mm at 40 AU.
            The surface density profile is defined by $\gamma=0.5$, $R_{tr} = 20$ AU, $\Sigma_{tr} = 8$ g/cm$^2$
            (solid lines), and $\Sigma_{tr} = 30$ g/cm$^2$ (dotted lines).}
   \label{fig:alpha_radius}
\end{figure}

We illustrate the effects of radial changes of the maximum grain size on the
disk  emission at millimeter waelengths by computing a small grid of models with different grain properties.
Specifically, we consider three different radial profiles of the maximum grain size 
$a_{\rm{max}}(R)$ across the disk, which are obtained by varying the power-law 
index $b_{\rm{max}}$ as shown in Figure~\ref{fig:amax_radius}. 
Since the spectral index of the dust opacity at mm-wavelengths $\beta$ depends 
on the maximum grain size of the emitting dust, a radial variation in $a_{\rm{max}}(R)$ 
corresponds to a radial variation in $\beta(R)$. 

Figure~\ref{fig:beta_radius} shows the radial profile of $\beta$ for the profiles of $a_{max}$ 
presented in Figure~\ref{fig:amax_radius}. The black line shows the case of a maximum grain size, 
which is constant throughout the disk, or $b_{\rm{max}} = 0$. Since $a_{\rm{max}}(R)$ does not 
change with the radius, $\beta(R)$ also has the same value in the disk. Red and blue lines show 
maximum grain size profiles, which decrease with radius, although with different power law indices 
as indicated in the two figures. For the dust model considered in this paper when $a_{\rm{max}} \gtrsim 0.05$ cm 
there is an anti-correlation between the maximum grain size and dust opacity spectral index, 
and this explains the general trend seen in Figures \ref{fig:amax_radius} and \ref{fig:beta_radius}. 
The bump seen at about $60 - 70$ AU in the $\beta(R)$ profile for the $b_{\rm{max}} = - 2.5$ case 
(blue line) is due to the increase in the dust opacity and its spectral index when 
$a_{\rm{max}} \approx \lambda/2\pi$ (see, e.g,  Ricci et al. 2010).

We apply the disk model described in the previous section to derive the disk 
surface brightness distribution at 1 and 3 mm and the radial profile of the spectral
index $\alpha_{\rm{1-3mm}}(R)$  for different radial profiles of the grain size distribution. 
To this end, we adopt the disk surface density 
parameters that best fit the interferometric data of CQ Tau:  $\gamma = 0.5$, $R_{tr} = 20$ AU, 
and $\Sigma_{tr} = 8$ (see Section \ref{sec:fitting}). 
We also consider a larger value for $\Sigma_{tr}$, i.e. 30 g/cm$^2$ to investigate the 
impact of a larger dust opacity on our analysis.

The top panel of Figure~\ref{fig:alpha_radius} shows the spectral index $\alpha_{\rm{1-3mm}}(R)$.
As a general trend, $\alpha_{\rm{1-3mm}}(R)$  follows the profile of $\beta(R)$. This is not surprising, 
since the difference $\alpha - \beta$ at a given radius  would be equal to 2 if the dust emission at these 
wavelengths is optically thin and in the Rayleigh-Jeans tail of the spectrum. 
If either one or both of these conditions is not strictly met, then $\alpha_{\rm{1-3mm}}$ would decrease, 
hence $\alpha - \beta < 2$. The lower panel of Figure~\ref{fig:alpha_radius} shows that 
this is the case at all radii for the models presented here.  In particular, the innermost regions of 
the disk are dense enough to make the dust emission depart from the optically thin limit 
at these wavelengths. In the outer regions of the disk midplane, the dust temperature is cold 
enough to make the disk emission deviate from the Rayleigh-Jeans approximation. The combination 
of these two effects explains why $\alpha - \beta$ is always lower than 2 for our disks; the largest deviation 
corresponds to the case of the denser disks (dotted lines) because of the higher optical depths.

The models discussed in this section have been computed for a fully flared disk. If the flaring angle is smaller due to dust sedimentation, disks would be cooler and departures from Rayleigh-Jeans regime would extend to smaller radii; a simple derivation of $\beta$ from $\alpha$ ($\beta\sim\alpha-2$) would  result in underestimating $\beta$ by large factors. In general, the disk geometry should be considered as a free parameter in the fitting procedure, unless additional constraints are available.

The examples of Figure~\ref{fig:alpha_radius} illustrate how the  radial profile of the spectral index 
$\alpha_{\rm{1-3mm}}$ is not uniquely determined by the dust properties, but depends on the 
absolute values and radial dependence of the surface density profile and temperature. 
This shows that an accurate determination of $\beta(R)$ cannot be obtained by $\alpha(R)$ alone.
Disentangling the dependence of $\beta$ on the radius from the surface density and temperature 
profiles can only be done by simultaneously fitting a set of spatially resolved data
at different sub-mm wavelengths, as we  discuss in the following sections.

\section{Model fitting to the visibility data of CQ Tau}
\label{sec:fitting}

In our models, the surface density profile is defined by three parameters
$(\gamma, R_{tr}, \Sigma_{tr}$; see Eq.~\ref{eq:selfsimilar_solution})
which can vary independently. Two additional free parameters 
($a_{\rm{max,0}}, b_{\rm{max}}$; see Eq.~\ref{eq:amax_PL})  describe the dust properties.
All the other quantities are either fixed, or computed self-consistently, as described in
Section \ref{sec:disk_model}.
We assume fully flared models because they are appropriate for CQ Tau, as shown by its SED (Natta et al.~2001).
We extended the procedure developed
by Banzatti et al.~(2011) to simultaneously fit interferometric visibilities at multiple wavelengths. 
A brief summary of the procedure is as follows: For each set of the disk model free parameters, we produce
a theoretical image of the disk at each wavelength $\lambda_i$. 
This image is Fourier transformed and sampled at the appropriate positions on the ($u,v$) plane corresponding
to the observed samples. We then compute the $\chi_{\lambda_i}^2$ value at each wavelength as in Banzatti et al.~(2011).
The global value of the $\chi^2$ for the parameter set is computed as the sum of the $\chi_{\lambda_i}^2$ 
over the observed $\lambda_i$. To find the best fitting model, we searched the minimum of the
$\chi_{sum}^2$ as a function of all parameters.

For our analysis, we use the PdBI datasets at 1.3 and 2.7~mm and the VLA dataset at 
7~mm presented in Banzatti et al. (2011). Data for CQ Tau at longer wavelengths do not have enough sensitivity (1.3~cm) 
or are dominated by
ionized gas emission (3.6~cm). The SMA data at shorter wavelengths (0.87~mm) were excluded from the fit for 
various reasons: The dataset is very limited both in terms of the number of sampled points in the ($u,v$) plane 
and of the range 
of baselines probed; in addition, the fits from Banzatti et al.~(2011) showed that the amplitude scale 
for these data may be incorrect. This latter point is very clear when looking at Figure~4 in Banzatti et al.~(2011):
While the surface density normalization is consistent at the other wavelengths, the fits at 0.87~mm require 
a normalization larger by a factor $\sim$2--3. This could be caused by an incorrect flux calibration of
the  SMA observations. As we cannot be sure of the cause for this and cannot properly correct the 
data, we prefer to omit them in our analysis.

\begin{figure}[]
   \centering
   \includegraphics[width=\columnwidth]{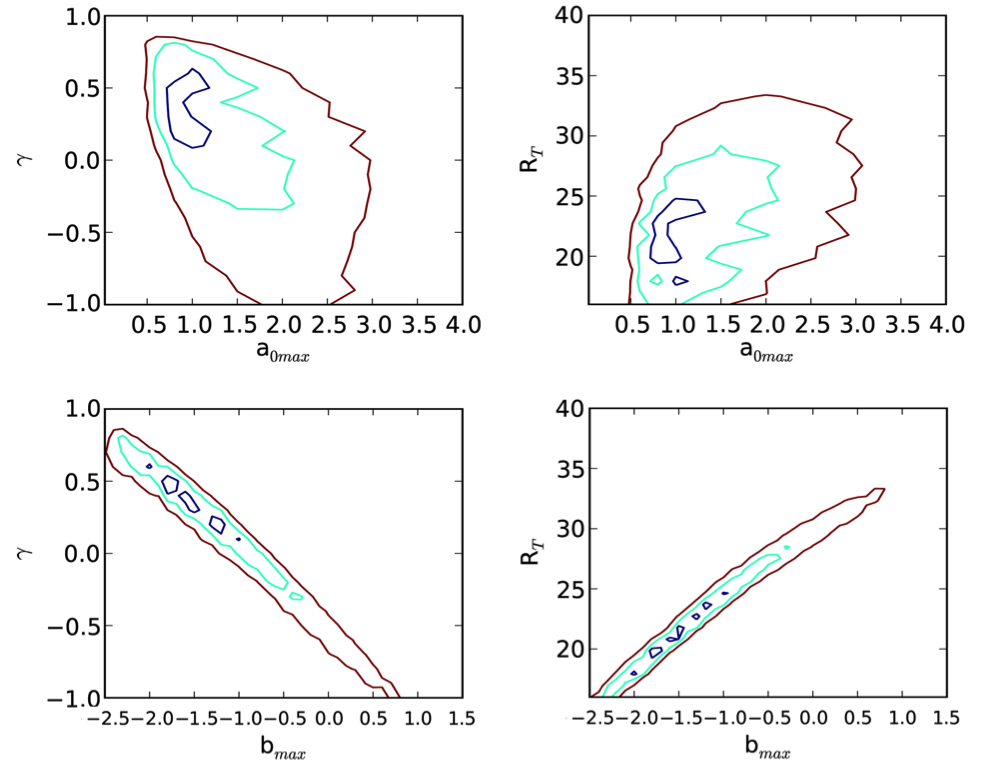}
   \caption{$\chi_{sum}^2$ hypercube projections on each coordinate (first row) and on different planes (second and third rows).
               To produce each plot we have chosen the value of the other parameters corresponding
               to the minimum of $\chi_{sum}^2$ at each position on the planes shown.
               The blue, green and red lines correspond respectively to the $1$, $2$ and $3 \sigma$
               (i.e. $68.3 \%$, $95.4 \%$ and $99.7 \%$) confidence levels.}
   \label{fig:cqtau_fit_2d}
\end{figure}

\begin{table}[]
   \centering
   \caption{Values of the best fit model parameters}
   \begin{tabular}{l c  c  c  c  c  c  c} \hline\hline

      \centering
      &$\gamma$             &  $\Sigma_{tr}~[\rm{g/cm^2}]$ &  $R_{tr}~[\rm{AU}]$          &  $b_{\rm{max}}$             &    $a_{\rm{max,0}}~[\rm{cm}]$      \\
      \hline
$^1$)      &$0.5_{-0.3}^{+0.1}$  &  $8_{-2}^{+2}$      &  $20_{-3}^{+4}$     &  $-1.8_{-0.2}^{+0.6}$  &  $0.8_{-0.1}^{+0.4}$  \\
      \hline
$^2$)      &$-0.5$               &  $5.0$                 &  $30.0$                &  $0.0$                 &  $1.5$                \\
      \hline
   \end{tabular}
      
$^1$) Model parameters and corrensponding confidence levels at $1\sigma$. 
$^2$) Same as above for the case of a radially constant dust opacity (i.e. $b_{\rm{max}}=0$).
   \label{tab:bestfit_model}
\end{table}

As for the disk inclination $i$ and position angle $PA$ (measured east of north) 
of the CQ Tau disk, we took the values of $i=34^\circ$ and $PA=42^\circ$, as constrained from the 
molecular gas observations of Chapillon et al. (2008).

\section{Results}
\label{sec:results}

The values of the best fit model parameters obtained by our analysis are listed 
in Table~\ref{tab:bestfit_model}. For each parameter we report the 1$\sigma$ confidence 
level intervals, following the analysis 
of the $\Delta\chi^2$ projections (see Press et al.~2007).
In the same Table, we also list the values of the best fit model parameters obtained when 
we force the same dust properties throughout the disk (i.e. $b_{\rm{max}}=0$). 
The best fit model shows a decrease the maximum grain size with radius
with a grain size distribution characterized by $b_{\rm{max}} \approx -1.8$ and $a_{\rm{max,0}} 
\approx 0.8$ cm. This corresponds to a disk populated by dust grains, which are grown to 
sizes larger than $\approx$ 1~cm in the inner regions ($R < 40$ AU) and as large as a few 
mm in the outer regions (see top panel in Figure~\ref{fig:ramax_bestfit}).

In figure~\ref{fig:cqtau_fit_2d} we show the projections of the $\chi^2$ hypercube 
on the 2D planes that include the grain size distribution parameters $a_{\rm{max,0}}$
and $b_{\rm{max}}$ and the surface density parameters, $\gamma$ and $R_t$. 
The figure shows that the disk surface density profile is strongly correlated 
with the radial profile of the grain size distribution. More specifically, 
the values of $\gamma$ derived, assuming constant dust opacity, 
$b_{max}=0$, are systematically lower  than those obtained when 
the radial variation in the grain size distribution is considered. 
We argue that this correlation between the surface density and the 
grain size distribution might explain some of the low values of $\gamma$ 
derived under the assumption of constant dust opacity by modeling single wavelength 
millimeter-wave observations (see, e.g., Isella et al. 2009). 

\begin{figure}[]
   \centering
   \includegraphics[width=7.5cm]{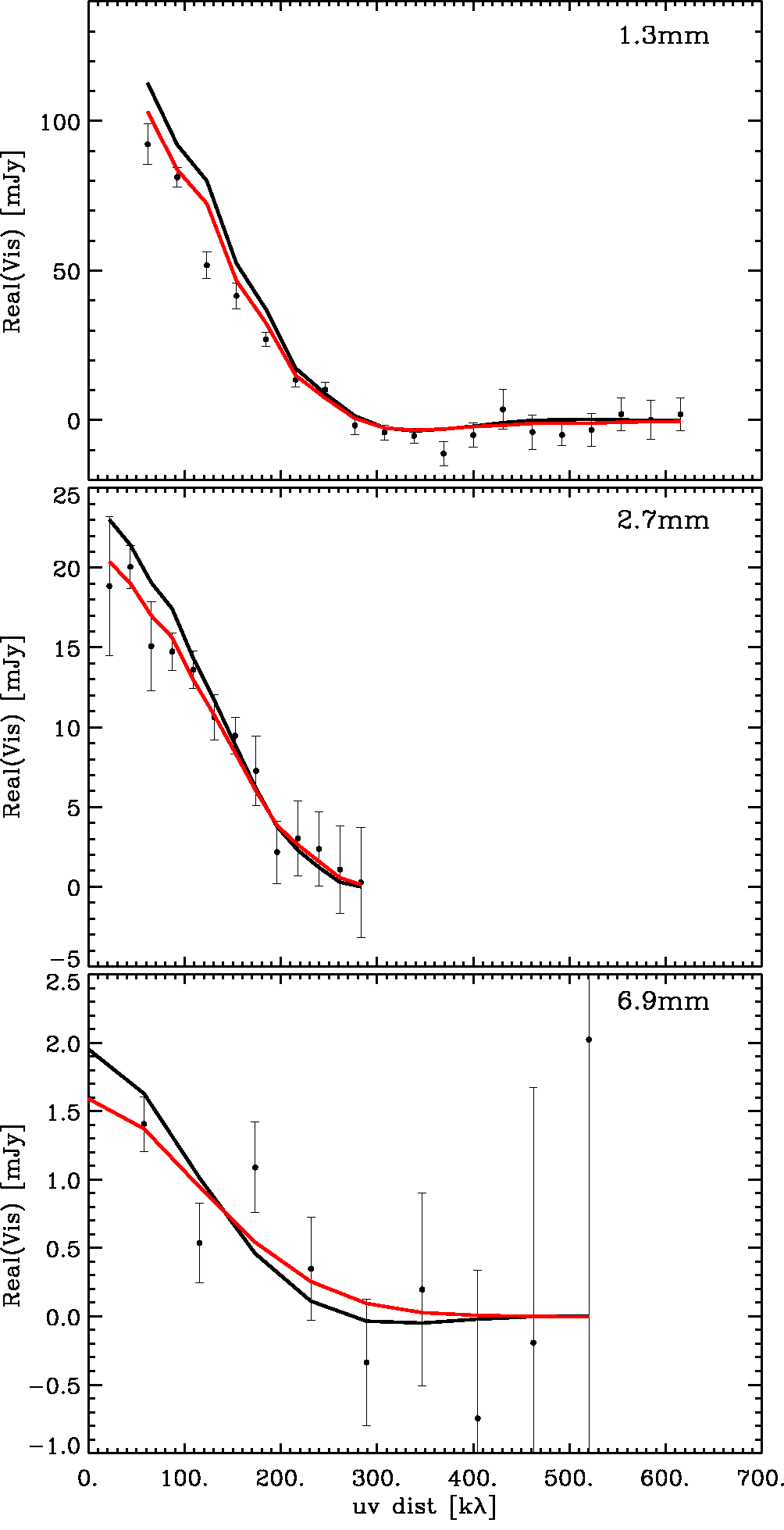}
   \caption{Comparison between the data (dots) and the best-fit model predictions (solid lines)
            of the real part of the interferometric visibilities as a function of deprojected baseline length.
            The solid lines show the predicted visibilities for the best-fit model (red lines)
            and the best-fit model with $b_{\rm{max}}=0$ (black line).}
   \label{fig:visib_obs_mod}
\end{figure}

The comparison between the observed visibilities and the results of our model fits
are shown in 
Fig.~\ref{fig:visib_obs_mod}.
To correct for the disk inclination, we deprojected the baselines for the assumed inclination
($i=34^\circ$) and position angle ($PA=42^\circ$) of the CQ Tau disk.
The real part of the complex visibilities has then been averaged in concentric 
annuli of deprojected $u,v$ distances from the disk center 
(in the figure, visibilities have been binned with widths of $\sim$30, 20, and $58 k\lambda$ 
for data at $1.3$, $3$, and $7$~mm, respectively).
The black lines represent the best-fit model predictions obtained in the case 
that we assume the same dust properties throughout the disk, when $b_{\rm{max}}=0$.
This model lies inside the $3\sigma$ confidence region, and although it does 
not correspond to the best fit, it is not rejected by our analysis. 

\begin{figure}[tbp]
   \centering
   \includegraphics[width=8.cm]{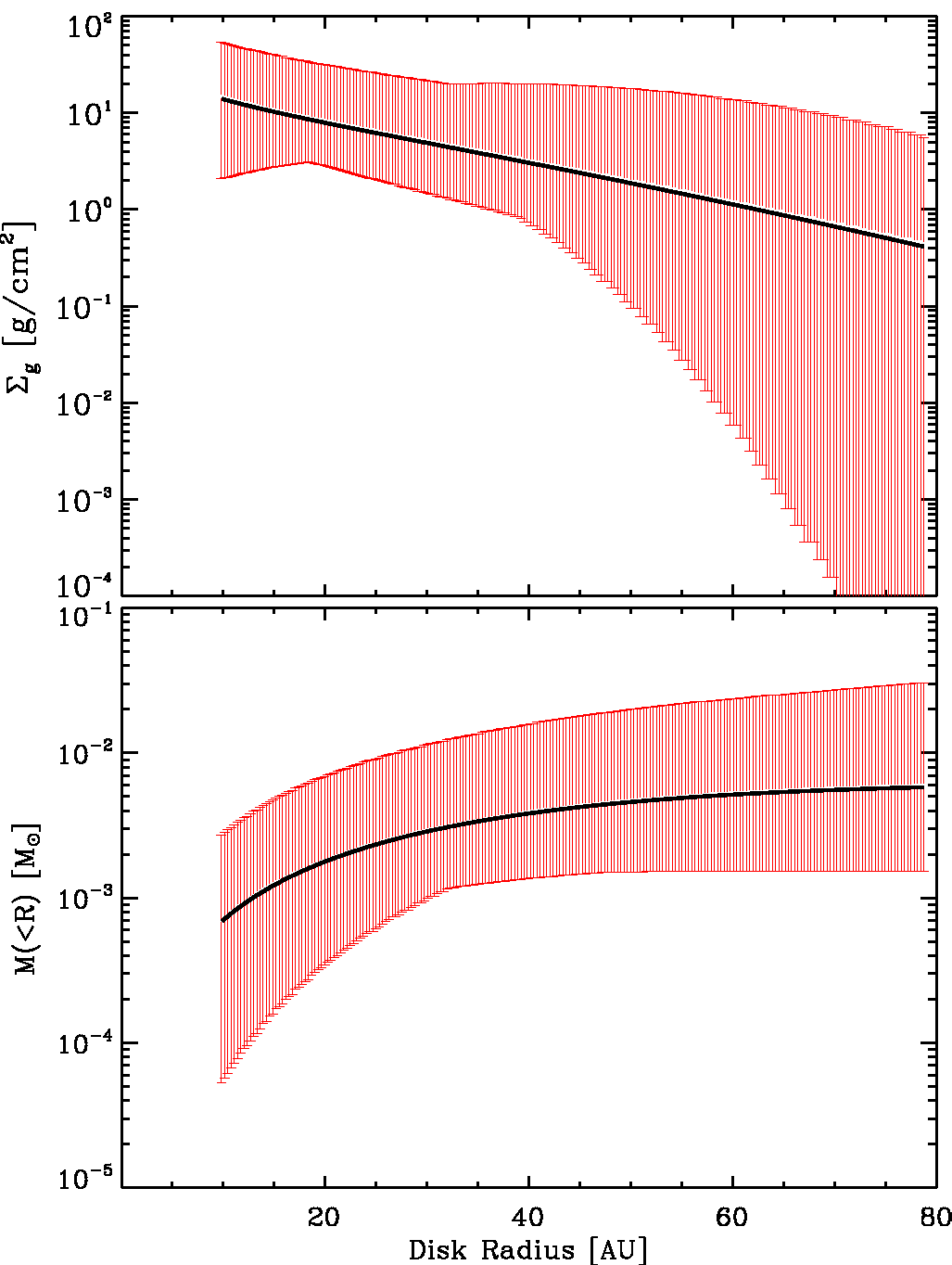}
   \caption{Gas surface density (\textit{top panel}) and cumulative mass (\textit{bottom panel}) profiles
            as a function of the radius for the best-fit model.
            The red hatched areas indicate the allowed range of values within the $3 \sigma$ uncertainty range.}
   \label{fig:rsigma_bestfit}
\end{figure}

The resulting surface density and cumulative mass radial profiles of the best fit model 
(first row in Table~\ref{tab:bestfit_model}) are shown in the top and bottom panels of 
Figure~\ref{fig:rsigma_bestfit}, respectively. These panels show disk radii up to about 80~AU. 
This is because the disk density at larger distances from the central star gets so low that the 
disk becomes optically thin to the stellar radiation, and no temperature structure can be 
calculated using the two-layer approximation. However, the lower panel of Figure~\ref{fig:rsigma_bestfit} 
shows that the radial profile of the cumulative mass distribution of the best fit model 
is already very close to being flat at $\approx 80$~AU, indicating that the amount of disk 
mass lost because of this disk truncation is negligible for our analysis. 
The total disk gas mass, assuming a gas-to-dust mass ratio of 100, is  $M_g \approx 0.006~M_{\odot}$.

When looking at the results of the model fitting and their uncertainties, one should 
keep in mind the strong correlation shown in Figure~\ref{fig:cqtau_fit_2d}.
As a consequence, some of the model free parameters cannot vary independently within 
the uncertainty intervals given in Table 1, and some disk properties are better constrained 
than it may appear. For example, if the maximum grain size is constant with 
radius ($b_{\rm{max}}=0$), the surface density profile required to have a good fit of 
the data is characterized by $\gamma = - 0.5$, $R_{tr} = 30$ AU, and
$\Sigma_{tr} = 5$ g/cm$^2$, which corresponds to a disk gas mass of 
$M_g \sim 0.008~M_{\odot}$. This value is only 30\% higher than our best fit value of 0.006 $M_\odot$.


\section{Discussion}
\label{sec:discussion}

\begin{figure}[tbp]
   \centering
   \includegraphics[width=8.cm]{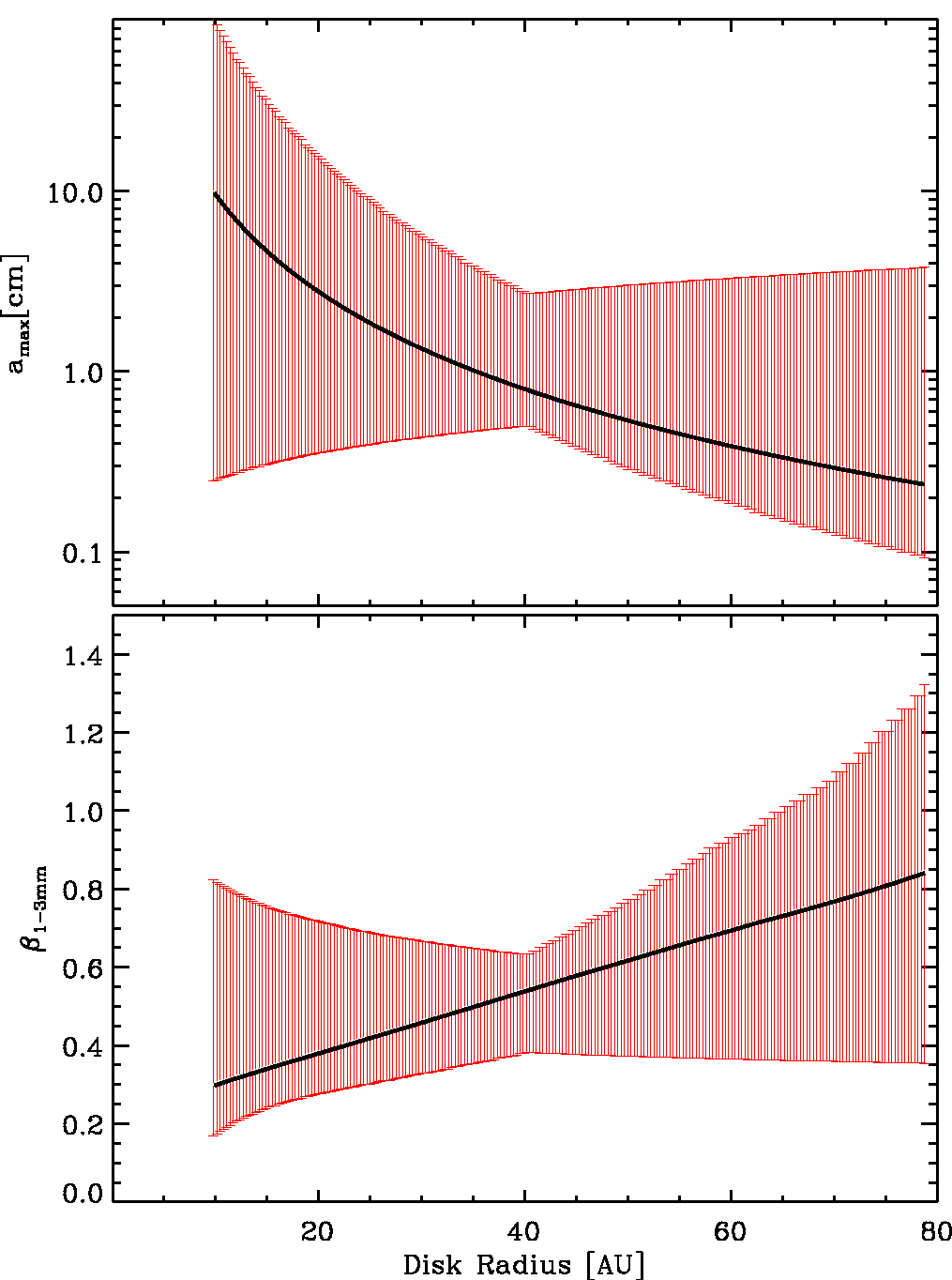}
   \caption{Maximum grain size (\textit{top panel}) and slope of the millimeter dust opacity (\textit{bottom panel})
            as function of the radius for the best-fit model.
            The red hatched areas indicate the allowed range of values within the $3 \sigma$ uncertainty range.}
   \label{fig:ramax_bestfit}
\end{figure}

The results presented in Section~\ref{sec:results} indicate a moderate
radial dependence of the grain sizes in the disk around CQ Tau.
From the best fit values, the CQ Tau disk shows (see top panel in Figure~\ref{fig:ramax_bestfit})
the presence of centimeter-sized grains in the inner disk ($R < 40$ AU) and
mm-sized grains in the outer disk ($R > 40$ AU).
The value of $\beta$ increases from $\approx 0.4$ at about 20~AU to $\approx 0.8$ at 80~AU and always remains  lower than 1.
Dust has been reprocessed and has grown in size not only in the inner
denser regions but also in the outer disk.
Models with constant $\beta \sim 0.5$ are consistent with the data within the $3\sigma$ confidence interval.

The radial dependence of the dust properties in the CQ Tau disk obtained in this paper confirms the 
suggestion of Banzatti et al. (2011) that the inner regions of the CQ Tau disk contain
grains larger than those in the outer regions.
Unlike the analysis presented here, they  separately fitted each dataset from 0.87 mm to 7 mm using
a two-layer disk model with a radial constant dust opacity distribution and adopted
a power law $\Sigma(R) = \Sigma_0(R/R_0)^{p}$ profile for the surface density distribution.
They obtained a variation in the fitted power-law surface density distribution
with the slope $p$ which increases with wavelength from $\approx 0$ at 1.3 mm to $\approx 0.5$ at 7 mm.
Assuming that the unphysical variation  of the surface density profile of the disk is due to radial variations
in the dust opacity spectral index $\beta$ (Isella et al. 2010),
they concluded that $\beta(R)$ must increase from $\sim 0.0$ in the inner disk
to $\beta \sim 1.4$ at $R \approx 63$ AU. Our analysis, based on models that allow for 
radial variation in $\beta$, favors slightly lower values of $\beta$ in the outer disk. 
We think that this difference is likely due to the  weight that the two techniques 
give to the observations at the various wavelengths. The approach used by  Isella et al. (2010)
gives the same weight to the results obtained at different wavelengths, while the global fit 
adopted here weights each observation with its own uncertainty. It should also be noted
that the disk models themselves are slightly different in Banzatti et al.~(2011), where a
power law $\Sigma(r)$ profile was adopted. The quality of the datasets available for 
CQ~Tau does not warrant an extensive comparison of the two methods, which should be performed 
as soon as better datasets are available.

A radial dependence of $\beta$ from lower to higher values for increasing radii
has been found in all disks analyzed so far.
Guilloteau et al. (2011) characterized the dust radial distribution in a sample of $20$ disks
in the Taurus-Auriga star-forming region by combining observations at 1.3 and 3 mm with the IRAM PdBI
interferometer.
Unlike this study, they fitted the data with parametric disk models
and found a radial dependence of  $\beta$, which appears to typically increase from low values
($\sim 0-0.5$) within $\sim 50-100$ AU from the central star to about $1.5-2$ at the disk 
outermost regions. P\'erez et al. (2012) investigated the radial variation in the dust opacity 
across the AS 209 circumstellar disk by combining spatially resolved observations from 
sub-millimeter to centimeter wavelengths; they used a two-layer disk model similar to the one 
described in this paper and followed the procedure outlined in Isella et al. (2010) and Banzatti et al. (2011).
Their results indicate that $\beta(R)$ increases from $\beta < 0.5$ at $\approx 20$~AU to
$\beta > 1.5$ for $R \gtrsim 80$~AU.

This general trend of decreasing maximum grain size with increasing radius
is predicted by theoretical models of dust evolution in circumstellar gasous disks.
According to simple theoretical considerations, the efficiency of the radial drift of solid particles 
depends on their coupling with the gas and large grains are expected to drift inward more 
rapidly than small grains (Whipple 1972; Weidenschilling 1977). Hence, one naturally expects 
large grains in the inner region and small grains in the outer regions.
If this drag is as efficient as the laminar theory predicts, dust particles of few mm in size would be removed
from the outer disk in a much shorter timescale than the typical age of protoplanetary disks of few million years (Brauer et al. 2007).
Recent dust evolution models, which consider the effects of radial drift, turbulent mixing, coagulation, and fragmentation,
have been performed by Brauer et al. (2008) and further expanded by Birnstiel et al. (2010a) to include the viscous evolution of the gas disk.
Their models predict a radial dependence of the grain growth level that is a function of the local physical conditions in the disk
(temperature, density, strength of the turbulence, and fragmentation velocity).
However, they show that the combination of radial drift and fragmentation is a strong limitation to particle growth and
represents an obstacle for the formation of planetesimals (known as the \textit{fragmentation barrier}).
Assuming that the radial drift is halted by some unknown physical mechanism, Birnstiel et al (2010b) compared the 
predictions of dust evolution models, including grain coagulation and fragmentation, with the results of sub-mm observations of disks
(Ricci et al. 2010a,b).  In most of the adopted models, they predicted steady-state grain size distributions that
lead to  dust opacity index $\beta(R)$ that increases with radius, which agree with our finding.

Within the general trend described above, models predict that the combined effect of the many
physical processes involved should produce grains with different properties in different disks.
This seems to be confirmed by the few observational results obtained so far,
which seem to indicate differences in the values of $\beta$ in the outer regions of different disks.
In particular, CQ Tau seems to have very evolved grains also in the outer disk.
However, one should keep in mind that the CQ Tau disk is small compared to others
and  that our results are  affected by  large uncertainties.

\section{Summary and conclusions}
\label{sec:conclusions}

We have attempted to constrain the radial distribution of the grain 
size distribution in the disk around CQ Tau. We propose a new method that is 
based on disk models that include the radial variations in the grain properties in 
a self-consistent way. These models are used to  simultaneously fit a set of
interferometric observations at different wavelengths and allow us to constrain 
at the same time the disk structure and the dust properties.

In CQ Tau, we find that the similarity solution for the viscous evolution of 
Keplerian disks (Lynden-Bell \& Pringle 1974) provides a good fit to the 
multi wavelength continuum observations at 1.3, 2.7, and 7 mm.
Although a model with a constant grain size distribution across the disk is still 
marginally consistent with the interferometric data, the model which best fits 
the data requires a radial dependence of the dust opacity. The resulting 
maximum grain size distribution is found to decrease with the radius 
from a few cm at $\sim 20 - 40$ AU from the star to a few mm further away from the star. 
Although the trend is similar to the prediction of dust grain evolution models, 
which include coagulation and fragmentation (Birnstiel et al. 2010b), it seems 
that grains are more evolved in the outer disk than in other objects investigated 
recently with other methods. This may be due to the large uncertainties present 
in all cases but may also reflect the relatively small size of the CQ Tau disk.

The results show that our self-consistent approach can be successfully
applied to the analysis of mm interferometric data.
The investigation of radial variations in the dust properties is still
strongly limited by the relatively poor angular resolution and sensitivity of
the current sub-mm facilities. Observations with higher angular resolution and
sensitivity are needed to place more stringent constraints on the radial
variation in the dust opacity. For these purposes, the ALMA and VLA
arrays will play a crucial role in the near future.

\begin{acknowledgements}
We thank Tillman Birnstiel for many useful discussions on grain growth 
in circumstellar disks. This research was partly supported by the Italian Space Agency (ASI) 
as part of the ASI-INAF agreement number I/005/11/0. A.I. acknowledge support from
NSF award AST-1109334.
\end{acknowledgements}


\begin{thebibliography}{}

   \bibitem[2005]{And05} Andrews, S. M., \& Williams, J. P. 2005, ApJ 631, 1134
   \bibitem[2011]{Arm11} Armitage 2011 ARA\&A, 49, 195A
   \bibitem[2011]{Banz11} Banzatti, A., Testi, L., Isella, A., et al. 2011, A\&A, 525, A12
   \bibitem[1991]{Beck91} Beckwith, S. V. W., \& Sargent, A. I. 1991, ApJ 381, 250
   \bibitem[2010]{Birn10a} Birnstiel, T., Dullemond, C. P., \& Brauer, F. 2010, A\&A, 513, 79
   \bibitem[2010]{Birn10b} Birnstiel, T., Ricci, L., Trotta, F., Dullemond, C.P., Natta, A., Testi, L., Dominik, C., Henning, T., Ormel, C.W., \& Zsom, A., 2010, A\&A, 516, 14
   \bibitem[2013]{Boe13} Boehler, Y., Dutrey, A., Guilloteau, S., \& Pietu, V. 2013, MNRAS, 431, 1573
   \bibitem[2011]{Bor11} Borucki, W. J., Koch, D. G., Basri, G., Batalha, N., et al. 2011, ApJ, 736, 19
   \bibitem[2008]{Brau08} Brauer, F., Dullemond, C. P., Henning, Th., 2008, A\&A, 480, 859
   \bibitem[2007]{Brau07} Brauer, F., Dullemond, C. P., Johansen, A., et al. 2007, A\&A, 469, 1169
   \bibitem[1935]{Brug35} Bruggeman D.A.G., 1935, Ann. Phys. Leipzig 24, 636
   \bibitem[2008]{Chap08} Chapillon, E., Guilloteau, S., Dutrey, A., \& Pietu, V. 2008, A\&A, 488, 565   
   \bibitem[1997]{cg97} Chiang, E. I., Goldreich, P., 1997, \apj, 490,368
   \bibitem[1997]{cg01} Chiang, E. I., Joung, M. K., Creech-Eakman, M. J., et al. 2001, ApJ, 547, 1077
   \bibitem[2006]{Dra06} Draine, B. T. 2006, ApJ 636, 1114
   \bibitem[2001]{Dull01} Dullemond, C.P., Dominik, C., \& Natta, A. 2001, ApJ, 560, 957
   \bibitem[2004]{Dull04} Dullemond, C.P., Dominik, C., 2004, A\&A, 421, 1075
   \bibitem[2011]{Gui11} Guilloteau, S., Dutrey, A., Pietu, V., \& Boehler, Y. 2011, A\&A 529, 105
   \bibitem[2008]{Hart08} Hartmann, L., Calvet, N., Gullbring, E., D'Alessio, P., 1998 ApJ 495 385H
   \bibitem[2009]{Ise09} Isella, A., Carpenter, J. M., \& Sargent, A. I. 2009, ApJ 701, 260
   \bibitem[2010]{Ise10} Isella, A., Carpenter, J. M., \& Sargent, A. I. 2010, ApJ 714, 1746 
   \bibitem[2005]{Johan05} Johansen A., Klahr, H. 2005, ApJ, 634, 1353J
   \bibitem[1993]{Lis93} Lissauer, J. J. 1993, ARA\&A, 31, 129
   \bibitem[2009]{Lom09} Lommen, D., Maddison, S. T., Wright, C. M., van Dishoeck, E. F., Wilner, D. J., Bourke, T. L. 2009, A\&A, 495, 869
   \bibitem[2007]{Lom07} Lommen, D., Wright, C. M., Maddison, S. T., Joergensen, J. K., Bourke, T. L., van Dishoeck, E. F., Hughes, A., Wilner, D. J., Burton, M., van Langevelde, H. J. 2007, A\&A, 462, 211
   \bibitem[1974]{Lyn74} Lynden-Bell, D., \& Pringle, J. E. 1974, MNRAS, 168, 603
   \bibitem[2000]{Natta00} Natta, A., Grinin, V., \& Mannings, V. 2000, Protostars and Planets IV, 559
   \bibitem[2001]{Natta01} Natta, A., Prusti, T., Neri, R., Wooden, D., Grinin, V., \& Mannings, V. 2001, A\&A, 371, 186
   \bibitem[2004]{Natta04} Natta, A., Testi, L., Neri, R., Shepherd, D. S., \& Wilner, D. J. 2004,
                           A\&A 416, 179
   \bibitem[2007]{Natta07} Natta, A., Testi, L., Calvet, N., Henning, Th., Waters, R., \& Wilner,
                           D., in Reipurth, B., Jewitt, D., Keil, K. (eds.), Protostars \&
                           Planets V. University of Arizona Press. Tucson. 2007. p. 783
   \bibitem[2012]{Perez12} Perez, Laura M., Carpenter, John M., Chandler, Claire J., Isella, Andrea,
                          Andrews, Sean M., Ricci, Luca, Calvet, Nuria, Corder, Stuartt A., Deller, Adam T.,
                          Dullemond, Cornelis P., Greaves, Jane S., Harris, Robert J., Henning, Thomas,
                          Kwon, Woojin, Lazio, Joseph, Linz, Hendrik, Mundy, Lee G., Sargent, Anneila I.,
                          Storm, Shaye, Testi, Leonardo, Wilner, David J. 2012, ApJ 760L, 17P
   \bibitem[1974]{Per97} Perryman, M. A. C., Lindegren, L., Kovalevsky, J., et al. 1997, A\&A, 323, L49
   \bibitem[2007]{P07} Press W.H., Teukolsky S.A., Vetterling W.T., and Flannery B.P., {\it Numerical Recipes: The Art of Scientific Computing}, Third Edition (New York: Cambridge University Press, 2007)
   \bibitem[2011]{Ricci11} 	Ricci, L., Mann, R. K., Testi, L., Williams, J. P., Isella, A., Robberto, M., Natta, A., Brooks, K. J. 2011, A\&A, 525, 81
   \bibitem[2010]{Ricci10a} Ricci, L.; Testi, L.; Natta, A.; Neri, R.; Cabrit, S.; Herczeg, G.J. 2010a, A\&A, 512, 15
   \bibitem[2010]{Ricci10b} Ricci, L.; Testi, L.; Natta, A.; Brooks, K. J. 2010b, A\&A, 521, 66
   \bibitem[2006]{Rodm06} Rodmann, J., Henning, T., Chandler, C. J., Mundy, L. G., \& Wilner, D. J. 2006, A\&A 446, 211
   \bibitem[2003]{Sem03} Semenov, D., Henning, T., Helling, C., Ilgner, M., \& Sedlmayr, E. 2003, A\&A, 410, 611
   \bibitem[1973]{SS73}  Shakura, N. I., Syunyaev, R. A., 1973, \aap, 24, 337
   \bibitem[2003]{Tes03} Testi, L., Natta, A., Shepherd, D. S., \& Wilner, D. J. 2003, A\&A 403, 323
   \bibitem[2001]{Tes01} Testi, L., Natta, A., Shepherd, D. S., \& Wilner, D. J. 2001, ApJ, 554, 1087
   \bibitem[2012]{Ubach12} Ubach, C., Maddison, S. T., Wright, C. M., Wilner, D. J., Lommen, D. J. P., Koribalski, B. 2012, MNRAS, 425, 3137
   \bibitem[1984]{War84} Warren, S. G. 1984, ApOpt 23, 1206
   \bibitem[1984]{Weid77} Weidenschilling, S. J. 1977, MNRAS, 180, 57
   \bibitem[2001]{Wei01} Weingartner, J. C., \& Draine, B. T. 2001, ApJ 548, 296
   \bibitem[2001]{Whip72} Whipple, F. L. 1972, in From Plasma to Planet (New York: Wiley Interscience Division), 211
   \bibitem[2001]{Wil01} Wilner, D. J., Ho, P. T. P., Kastner, J. H., \& Rodríguez, L. F. 2000, ApJ, 534, L101
   \bibitem[2005]{Wil05} Wilner, D. J., D'Alessio, P., Calvet, N., et al. 2005, ApJ Letters 626, 109
   \bibitem[1996]{Zub96} Zubko, V. G., Mennella, V., Colangeli, L., \& Bussoletti, E. 1996, MNRAS, 282, 1321

\end{thebibliography}
\end{document}